\documentclass[aps,pra,showpacs,twocolumn,floatfix]{revtex4}
\usepackage{multirow}
\usepackage{bm,dcolumn,amsmath,graphicx}

\begin{document}

\title {Transitions between  the
$4f$-core-excited states in Ir$^{16+}$, Ir$^{17+}$, and Ir$^{18+}$
 ions for clock applications}

\author{U. I. Safronova}
\affiliation{Physics Department, University of Nevada, Reno,
Nevada 89557}
\author{V. V. Flambaum}
\affiliation {University of New South Wales, Sydney 2052, Australia}

\author{M. S. Safronova}
\affiliation{Department of Physics and Astronomy, University of
Delaware, Newark, Delaware 19716}
\affiliation{Joint Quantum Institute, NIST and the University of Maryland, College Park, Maryland 20899, USA}

\begin{abstract}
Iridium ions near $4f$-$5s$ level crossings are the leading candidates for a new type of atomic clocks with a  high
 projected accuracy and a very high sensitivity to the temporal variation of the fine structure constant $\alpha$.
  To identify spectra of these ions in experiment  accurate calculations of the spectra and electromagnetic
   transition probabilities should be performed.
Properties of the $4f$-core-excited states in Ir$^{16+}$,
Ir$^{17+}$, and Ir$^{18+}$
 ions are evaluated using  relativistic many-body perturbation
 theory and
Hartree-Fock-Relativistic method (COWAN code). We evaluate
excitation energies, wavelengths, oscillator strengths, and
transition rates. Our large-scale calculations includes the
following set of configurations: $4f^{14-k}5s^{m}5p^{n}$ with
$(k+m+n)$ equal to 3, 2, and 1 for the Ir$^{16+}$, Ir$^{17+}$, and
Ir$^{18+}$ ions, respectively. The $5s-5p$ transitions are
illustrated by the synthetic spectra in the 180 - 200~\AA ~range.
Large contributions   of magnetic-dipole
transitions to lifetimes of low-lying states in the
region below 2.5~Ry are demonstrated.

\end{abstract}
 \maketitle

\section {Introduction}

Selected transitions involving electron holes, i.e. vacancies in otherwise filled shells of
 atomic systems in highly-charged ions were shown to have frequencies within the range of optical
atomic clocks, have small systematic errors in the frequency measurements and be highly sensitive
 to the temporal variation of the fine structure constant $\alpha$ ~\cite{prl-11}.
Sympathetic cooling of highly-charged ions has been demonstrated in \cite{SchVerSch15}.

This work is motivated by these applications of highly-charged ions and recent experimental
 work including identification of
M1 transitions in
Ir$^{17+}$ spectra ~\cite{SchVerSch15,prl-15,PRB15}. In 2015, identification of the predicted $5s-4f$ level crossing
optical transitions was presented by Windberger{\it et al.\/}
\cite{prl-15}. The spectra of Nd-like W, Re,
Os, Ir, and Pt ions of particular interest for tests of
fine-structure constant variation  were explored \cite{prl-15}.
The authors exploited characteristic energy scalings to identify  the strongest
lines, confirmed the predicted $5s-4f$ level crossing, and benchmarked
advanced calculations.
\begin{figure}[h]
\centerline{ \includegraphics[scale=0.35]{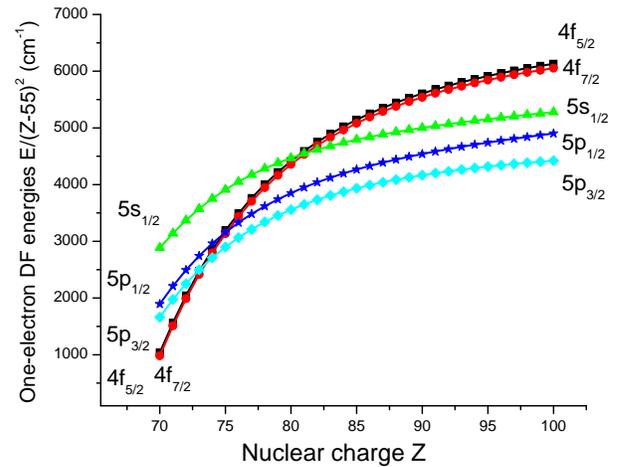}}
 \caption{Dirac-Fock binding energies of the $4f$ , $5s$, and $5p$ orbitals as a function of Z in Er-like ions.
 Er is a  rare earth element with $Z=68$.
 } \label{fig-hole}
\end{figure}

\begin{table*}
\caption{
 Energies (in 1000~cm$^{-1}$) in Pm-like Ir$^{16+}$, Nd-like Ir$^{17+}$, and Pr-like Ir$^{18+}$ ions given relative to
 the $4f^{13}5s^2\ ^2F_{7/2}$,  $4f^{13}5s\ ^3F_{4}$, and $4f^{13}\ ^2F_{7/2}$ ground states, respectively. }
\begin{ruledtabular}
\begin{tabular}{llrllrllr}
\multicolumn{1}{c}{Conf.}& \multicolumn{1}{c}{Level}&
\multicolumn{1}{c}{Energy}& \multicolumn{1}{c}{Conf.}&
\multicolumn{1}{c}{Level}& \multicolumn{1}{c}{Energy}&
\multicolumn{1}{c}{Conf.}& \multicolumn{1}{c}{Level}&
\multicolumn{1}{c}{Energy}\\\hline
\multicolumn{3}{c}{Pm-like
Ir$^{16+}$ ion}& \multicolumn{3}{c}{Nd-like Ir$^{17+}$ ion}&
\multicolumn{3}{c}{Pr-like Ir$^{18+}$ ion}\\
  $4f^{13}5s^2    $&$(^2F) ^2F _{7/2}$&    0.000&      $4f^{13}5s     $&$(^2F) ^3F  _{4}$&      0.000   &$4f^{13}    $&$(^2F) ^2F _{7/2}$&     0.000\\
  $4f^{13}5s^2    $&$(^2F) ^2F _{5/2}$&   25.909&      $4f^{13}5s     $&$(^2F) ^3F  _{3}$&      4.236   &$4f^{13}    $&$(^2F) ^2F _{5/2}$&    26.442\\
  $4f^{14}5s      $&$(^1S) ^2S _{1/2}$&   28.350&      $4f^{14}       $&$(^1S) ^1S  _{0}$&      5.091   &$4f^{12}5s  $&$(^3H) ^4H_{13/2}$&    60.142 \\
  $4f^{13}5s5p    $&$(^2F) ^4D _{7/2}$&  267.942&      $4f^{13}5s     $&$(^2F) ^3F  _{2}$&     26.174   &$4f^{12}5s  $&$(^3H) ^4H_{11/2}$&    67.687\\
  $4f^{12}5s^25p  $&$(^3H) ^4G_{11/2}$&  276.214&      $4f^{13}5s     $&$(^2F) ^1F  _{3}$&     30.606   &$4f^{12}5s  $&$(^3F) ^4F _{9/2}$&    70.096\\
  $4f^{12}5s^25p  $&$(^3H) ^2I_{13/2}$&  280.511&      $4f^{12}5s^2   $&$(^3H) ^3H  _{6}$&     33.856   &$4f^{12}5s  $&$(^1G) ^2G _{7/2}$&    74.664\\
  $4f^{12}5s^25p  $&$(^3F) ^4D _{7/2}$&  285.301&      $4f^{12}5s^2   $&$(^3F) ^3F  _{4}$&     42.199   &$4f^{12}5s  $&$(^3H) ^4H _{9/2}$&    87.749\\
  $4f^{13}5s5p    $&$(^2F) ^4D _{7/2}$&  286.286&      $4f^{12}5s^2   $&$(^3H) ^3H  _{5}$&     58.261   &$4f^{12}5s  $&$(^3H) ^2H_{11/2}$&    89.859\\
  $4f^{12}5s^25p  $&$(^1G) ^2H _{9/2}$&  287.909&      $4f^{12}5s^2   $&$(^3F) ^3F  _{2}$&     63.696   &$4f^{12}5s  $&$(^3F) ^4F _{3/2}$&    92.406\\
  $4f^{13}5s5p    $&$(^2F) ^4G _{9/2}$&  288.623&      $4f^{12}5s^2   $&$(^3H) ^3H  _{4}$&     66.296   &$4f^{12}5s  $&$(^3F) ^4F _{5/2}$&    92.940\\
  $4f^{13}5s5p    $&$(^2F) ^4F _{5/2}$&  289.959&      $4f^{12}5s^2   $&$(^3F) ^3F  _{3}$&     68.886   &$4f^{12}5s  $&$(^3F) ^4F _{7/2}$&    94.508\\
  $4f^{13}5s5p    $&$(^2F) ^4G _{5/2}$&  297.128&      $4f^{12}5s^2   $&$(^3H) ^3H  _{4}$&     89.455   &$4f^{12}5s  $&$(^1G) ^2G _{9/2}$&    98.486\\
  $4f^{12}5s^25p  $&$(^3H) ^4I_{11/2}$&  303.063&      $4f^{12}5s^2   $&$(^3P) ^3P  _{2}$&     91.765   &$4f^{12}5s  $&$(^3F) ^2F _{7/2}$&   101.805\\
  $4f^{12}5s^25p  $&$(^3H) ^4G _{9/2}$&  303.664&      $4f^{12}5s^2   $&$(^3P) ^3P  _{0}$&    101.073   &$4f^{12}5s  $&$(^3F) ^2F _{5/2}$&   102.646\\
  $4f^{12}5s^25p  $&$(^1D) ^2D _{3/2}$&  307.339&      $4f^{12}5s^2   $&$(^1I) ^1I  _{6}$&    101.537   &$4f^{12}5s  $&$(^1G) ^2G _{7/2}$&   119.089\\
  $4f^{12}5s^25p  $&$(^3F) ^4G _{5/2}$&  307.948&      $4f^{12}5s^2   $&$(^3P) ^3P  _{1}$&    107.843   &$4f^{12}5s  $&$(^3P) ^4P _{5/2}$&   122.306\\
  $4f^{12}5s^25p  $&$(^3F) ^4G _{7/2}$&  309.803&      $4f^{12}5s^2   $&$(^1D) ^1D  _{2}$&    117.322   &$4f^{12}5s  $&$(^3H) ^2H _{9/2}$&   123.040\\
  $4f^{12}5s^25p  $&$(^3H) ^4I _{9/2}$&  312.724&      $4f^{12}5s^2   $&$(^1S) ^1S  _{0}$&    178.055   &$4f^{12}5s  $&$(^3F) ^4F _{3/2}$&   123.434\\
  $4f^{12}5s^25p  $&$(^3F) ^4F _{5/2}$&  313.871&      $4f^{12}5s5p   $&$(^3H) ^5G  _{6}$&    312.027   &$4f^{12}5s  $&$(^3P) ^4P _{1/2}$&   129.563\\
  $4f^{13}5s5p    $&$(^2F) ^2D^{a}_{5/2}$&315.471&     $4f^{13}5p     $&$(^2F) ^3D  _{3}$&    319.802   &$4f^{12}5s  $&$(^1I) ^2I_{13/2}$&   132.511\\
  $4f^{13}5s5p    $&$(^2F) ^2G^{a}_{7/2}$&  317.004&      $4f^{12}5s5p   $&$(^3F) ^5D  _{4}$&    321.722   &$4f^{12}5s  $&$(^1I) ^2I_{11/2}$&   132.716\\
  $4f^{12}5s^25p  $&$(^3F) ^4G _{7/2}$&  317.387&      $4f^{13}5p     $&$(^2F) ^3G  _{4}$&    322.623   &$4f^{12}5s  $&$(^3P) ^4P _{3/2}$&   136.659\\
  $4f^{13}5s5p    $&$(^2F) ^4F _{3/2}$&  319.123&      $4f^{12}5s5p   $&$(^3H) ^3I^{b}_{7}$&  333.465   &$4f^{12}5s  $&$(^3P) ^2P _{1/2}$&   145.476\\
  $4f^{12}5s^25p  $&$(^3H) ^4I _{9/2}$&  332.682&      $4f^{12}5s5p   $&$(^3H) ^5G  _{5}$&    333.531   &$4f^{12}5s  $&$(^1D) ^2D _{5/2}$&   147.492\\
  $4f^{12}5s^25p  $&$(^3P) ^4P _{3/2}$&  335.326&      $4f^{12}5s5p   $&$(^3H) ^5G  _{6}$&    333.649   &$4f^{12}5s  $&$(^3P) ^2P _{3/2}$&   152.098\\
  $4f^{12}5s^25p  $&$(^3F) ^4G _{5/2}$&  337.196&      $4f^{12}5s5p   $&$(^3H) ^5I  _{5}$&    340.286   &$4f^{11}5s^2$&$(^4I) ^4I_{15/2}$&   179.577\\
  $4f^{14}5p      $&$(^1S) ^2P _{1/2}$&  337.757&      $4f^{12}5s5p   $&$(^1G) ^3H  _{4}$&    342.157   &$4f^{11}5s^2$&$(^4F) ^4F _{9/2}$&   201.648\\
 \end{tabular}
\end{ruledtabular}
\label{tab-energy}
\end{table*}

In the present work, we have employed Hartree-Fock-Relativistic method (COWAN code) and
relativistic many-body perturbation theory (RMBPT)
to
study Ir$^{16+}$, Ir$^{17+}$, and Ir$^{18+}$ ions.
Excitation energies, wavelengths, transition rates,  energies of the
lower and upper level, lifetimes, and branching ratios  from M1
and E1 transitions in Nd-, Pm-, and Pr-like Ir ions are evaluated. We have used
our results to construct synthetic spectra for all three ions.
Our goal was to also to investigate the general structure and level distributions in these ions.
For example, we find that in the case of Ir$^{18+}$ ion, the energies of the
$4f^{13}$, $4f^{12}5s$, and $4f^{11}5s^2$ configurations are within a
relatively small interval  below 274768~cm$^{-1}$. The first level with $5p$
electron, $4f^{12}5p\ (^3H) ^4G_{11/2}$, lies substantially higher, at
387658~cm$^{-1}$ which significantly affects the lifetimes of the lower states.

 We also employed RMBPT method
 for simpler one-particle, $4f^{14}nl$, and particle-hole configurations of
Ir ions to compare with HFR results.

The values for low-lying levels are presented in the paper, much more extensive set of results
is given in the Supplemental Material~\cite{suppl}.

\section {Level crossings and $4f$ electrons in  highly-charge ions}
Detailed investigation of level crossings relevant to the design of optical atomic clocks with highly-charged ions
and search for $\alpha$-variation has been carried out in
\cite{BerDzuFla12}. Ir ions have been considered in \cite{prl-11}.
Below, we discuss level crossings in ions similar to the ones studies in this work.

Correlation and relativistic effects for
the $4f - nl$ and $5p - nl$ multipole transitions in Er-like
tungsten were investigated in Ref.~\cite{pra-11-er}. Wavelengths,
transition rates, and line strengths were calculated for the
multipole (E1, M1, E2, M2, and E3) transitions between the excited
[Cd]$4f^{13}5p^6nl$, [Cd]$4f^{14}5p^5nl$ configurations
 and the ground [Cd]$4f^{14}5p^6$ state in Er-like W$^{6+}$ ion
  ([Cd]=[Kr]$4d^{10}5s^2$) using the relativistic
many-body perturbation theory, including the Breit
interaction.

The binding energies of the
$4f$, $5p$, and $5s$ orbitals  in Er-like ions \cite{pra-11-er} calculated  in Dirac-Fock
 approximation as  function of nuclear charge $Z$  are plotted in Fig.~\ref{fig-hole}. For better
presentation, we scaled the energies with a factor of $(Z-55)^2$.
We find that the   $4f$ orbitals are more
tightly bound than the $5p$ and $5s$ orbitals at low
stages of ionization, while  the  $5p$ and $5s$ orbitals are more
tightly bound than the $4f$ orbitals for
highly ionized cases. This leads to crossing  of $4f$  and $5s$ and $5p$ levels for some $Z$ leading to interesting cases of optical or near-optical transitions in selected  highly-charged ions.
Large cancellation in binding energy values near the crossing (near $Z=74$ in the example of Fig.~\ref{fig-hole})
make accurate calculations of transition energies and line strengths  very difficult
\cite{pra-11-er}.

In Ref.~\cite{pra-13-sm}, Safronova {\it et al.\/} reported
results of {\it ab initio} calculation of excitation energies,
oscillator strengths, transition probabilities, and lifetimes in
Sm-like ions with nuclear charge $Z$ ranging from 74 to 100. Sm has $Z=62$. One
of the unique atomic properties of the samarium isoelectronic
sequence is that the ground state changes nine times starting from
the [Kr]$4d^{10}5s^25p^64f^66s^2\ ^1S_0$ level for neutral
samarium, Sm~I, and ending with  the [Kr]$4d^{10}4f^{14}5s^2\
 ^1S_0$ level for 12 times ionized tungsten, W$^{12+}$ \cite{KraRalRea11}.

\begin{table*}
\caption{\label{tab2} Wavelengths ($\lambda$ in \AA), weighted
oscillator strengths (gf), weighted transition rates ($gA_r$ in
1/s) for transitions in Pm-like Ir$^{16+}$,  Nd-like Ir$^{17+}$,
and Pr-like Ir$^{18+}$ ions. }
\begin{ruledtabular}
\begin{tabular}{llllllll}
\multicolumn{1}{l}{Conf.} & \multicolumn{1}{l}{Level} &
\multicolumn{1}{l}{Conf.} & \multicolumn{1}{l}{Level} &
\multicolumn{1}{l}{$\lambda$ in \AA} & \multicolumn{1}{l}{gf} &
\multicolumn{1}{l}{$gA_r$ in 1/s} \\
\hline
 \multicolumn{6}{c}{$4f^{12}5s^25p - 4f^{12}5s5p^2$ transitions in Pm-like Ir$^{16+}$ ion}  \\
  $4f^{12}5s^25p$&$(^1I) ^2K_{15/2}$&    $4f^{12}5s5p^2$&$(^1I) ^2K^{a}_{15/2}$&     187.3222&    12.1796&  2.315[12]\\
  $4f^{12}5s^25p$&$(^3H) ^4I_{15/2}$&    $4f^{12}5s5p^2$&$(^3H) ^4I^{a}_{15/2}$&     188.4095&     7.6206&  1.432[12]\\
  $4f^{12}5s^25p$&$(^3H) ^4I_{13/2}$&    $4f^{12}5s5p^2$&$(^3H) ^4I^{b}_{13/2}$&     189.4141&    10.2303&  1.902[12]\\
  $4f^{12}5s^25p$&$(^1I) ^2H_{11/2}$&    $4f^{12}5s5p^2$&$(^1I) ^2H^{a}_{11/2}$&     190.2284&     7.7758&  1.433[12]\\
  $4f^{12}5s^25p$&$(^1G) ^2H_{11/2}$&    $4f^{12}5s5p^2$&$(^3H) ^2I^{b}_{11/2}$&     190.2870&     7.6461&  1.408[12]\\
  $4f^{12}5s^25p$&$(^1I) ^2K_{13/2}$&    $4f^{12}5s5p^2$&$(^1I) ^2K^{a}_{13/2}$&     190.5154&     9.7403&  1.790[12]\\
  $4f^{12}5s^25p$&$(^1I) ^2I_{13/2}$&    $4f^{12}5s5p^2$&$(^1I) ^2H^{a}_{11/2}$&     191.9017&     6.8966&  1.249[12]\\
  $4f^{12}5s^25p$&$(^1D) ^2F_{7/2} $&    $4f^{12}5s5p^2$&$(^1D) ^2F^{a}_{7/2} $&     192.6116&     5.6848&  1.022[12]\\
  $4f^{12}5s^25p$&$(^3H) ^4G_{11/2}$&    $4f^{12}5s5p^2$&$(^3H) ^4G^{b}_{11/2}$&     192.9081&     5.5967&  1.003[12]\\
  $4f^{12}5s^25p$&$(^3H) ^4I_{11/2}$&    $4f^{12}5s5p^2$&$(^3H) ^2I^{b}_{11/2}$&     193.3102&     7.1596&  1.278[12]\\
   \multicolumn{6}{c}{$4f^{12}5s^2 - 4f^{12}5s5p$ and $4f^{12}5s5p - 4f^{12}5p^2$  transitions in Nd-like Ir$^{17+}$ ion}  \\
  $4f^{12}5s5p$&$  (^1I) ^3K_{8}   $&     $4f^{12}5p2  $&$(^1I) ^3K^{ }_{8}   $&     186.9139&     5.5800&  1.065[12]\\
  $4f^{12}5s^2$&$  (^1I) ^1I_{6}   $&     $4f^{12}5s5p $&$(^1I) ^1I^{ }_{6}   $&     187.4003&     8.8646&  1.684[12]\\
  $4f^{12}5s^2$&$  (^3H) ^3H_{6}   $&     $4f^{12}5s5p $&$(^3H) ^3H^{b}_{6}   $&     190.4460&     7.6415&  1.405[12]\\
  $4f^{12}5s^2$&$  (^3H) ^3H_{4}   $&     $4f^{12}5s5p $&$(^3H) ^3H^{b}_{4}   $&     191.2123&     5.6115&  1.024[12]\\
  $4f^{12}5s^2$&$  (^3H) ^3H_{5}   $&     $4f^{12}5s5p $&$(^3H) ^3I^{b}_{6}   $&     194.1681&     7.6418&  1.352[12]\\
  $4f^{12}5s^2$&$  (^3H) ^3H_{6}   $&     $4f^{12}5s5p $&$(^3H) ^3I^{a}_{7}   $&     194.6655&     9.2353&  1.625[12]\\
  $4f^{12}5s^2$&$  (^1I) ^1I_{6}   $&     $4f^{12}5s5p $&$(^1I) ^1K^{ }_{7}   $&     195.6155&     8.9236&  1.555[12]\\
  $4f^{12}5s5p$&$  (^1I) ^1I_{6}   $&     $4f^{12}5p2  $&$(^1I) ^1I^{b}_{6}   $&     201.1890&     7.8226&  1.289[12]\\
  $4f^{12}5s5p$&$  (^1I) ^1K_{7}   $&     $4f^{12}5p2  $&$(^1I) ^3K^{ }_{8}   $&     213.6590&     6.9882&  1.021[12]\\
   \multicolumn{6}{c}{$4f^{11}5s^2 - 4f^{11}5s5p$ transitions in Pr-like Ir$^{18+}$ ion}  \\
  $4f^{11}5s^2$&$  (^2F) ^2F^{1}_{7/2} $&   $4f^{11}5s5p$&$ (^2F) ^2F^{1} _{7/2} $&   182.4907&     5.6811 &  1.138[12]\\
  $4f^{11}5s^2$&$  (^2L) ^2L^{ }_{15/2}$&   $4f^{11}5s5p$&$ (^2L) ^2L^{b} _{15/2}$&   184.2146&    11.5687 &  2.274[12]\\
  $4f^{11}5s^2$&$  (^2L) ^2L^{ }_{17/2}$&   $4f^{11}5s5p$&$ (^2L) ^2L^{b} _{17/2}$&   184.3734&     8.2009 &  1.609[12]\\
  $4f^{11}5s^2$&$  (^2I) ^2I^{ }_{13/2}$&   $4f^{11}5s5p$&$ (^2I) ^2I^{b} _{13/2}$&   186.7641&     7.6961 &  1.472[12]\\
  $4f^{11}5s^2$&$  (^2G) ^2G^{2}_{7/2} $&   $4f^{11}5s5p$&$ (^2G) ^2H^{2} _{9/2} $&   187.6959&     5.6285 &  1.066[12]\\
  $4f^{11}5s^2$&$  (^4I) ^4I^{ }_{15/2}$&   $4f^{11}5s5p$&$ (^4I) ^4I^{b} _{15/2}$&   187.7941&    10.0169 &  1.894[12]\\
  $4f^{11}5s^2$&$  (^4I) ^4I^{ }_{13/2}$&   $4f^{11}5s5p$&$ (^4I) ^4I^{b} _{13/2}$&   187.9089&     7.4924 &  1.415[12]\\
  $4f^{11}5s^2$&$  (^2H) ^2H^{2}_{9/2} $&   $4f^{11}5s5p$&$ (^2H) ^2I^{2} _{11/2}$&   188.0677&     6.8954 &  1.300[12]\\
  $4f^{11}5s^2$&$  (^2H) ^2H^{1}_{11/2}$&   $4f^{11}5s5p$&$ (^2I) ^2K^{b} _{13/2}$&   188.1826&     6.6946 &  1.261[12]\\
  $4f^{11}5s^2$&$  (^4I) ^4I^{ }_{11/2}$&   $4f^{11}5s5p$&$ (^4I) ^4I^{b} _{11/2}$&   188.3409&     6.1020 &  1.147[12]\\
  $4f^{11}5s^2$&$  (^4G) ^4G^{ }_{9/2} $&   $4f^{11}5s5p$&$ (^4I) ^4K^{b} _{11/2}$&   188.5680&     5.3585 &  1.005[12]\\
  \end{tabular}
\end{ruledtabular}
\end{table*}

\begin{figure*}[tbp]
\centerline{\includegraphics[scale=0.35]{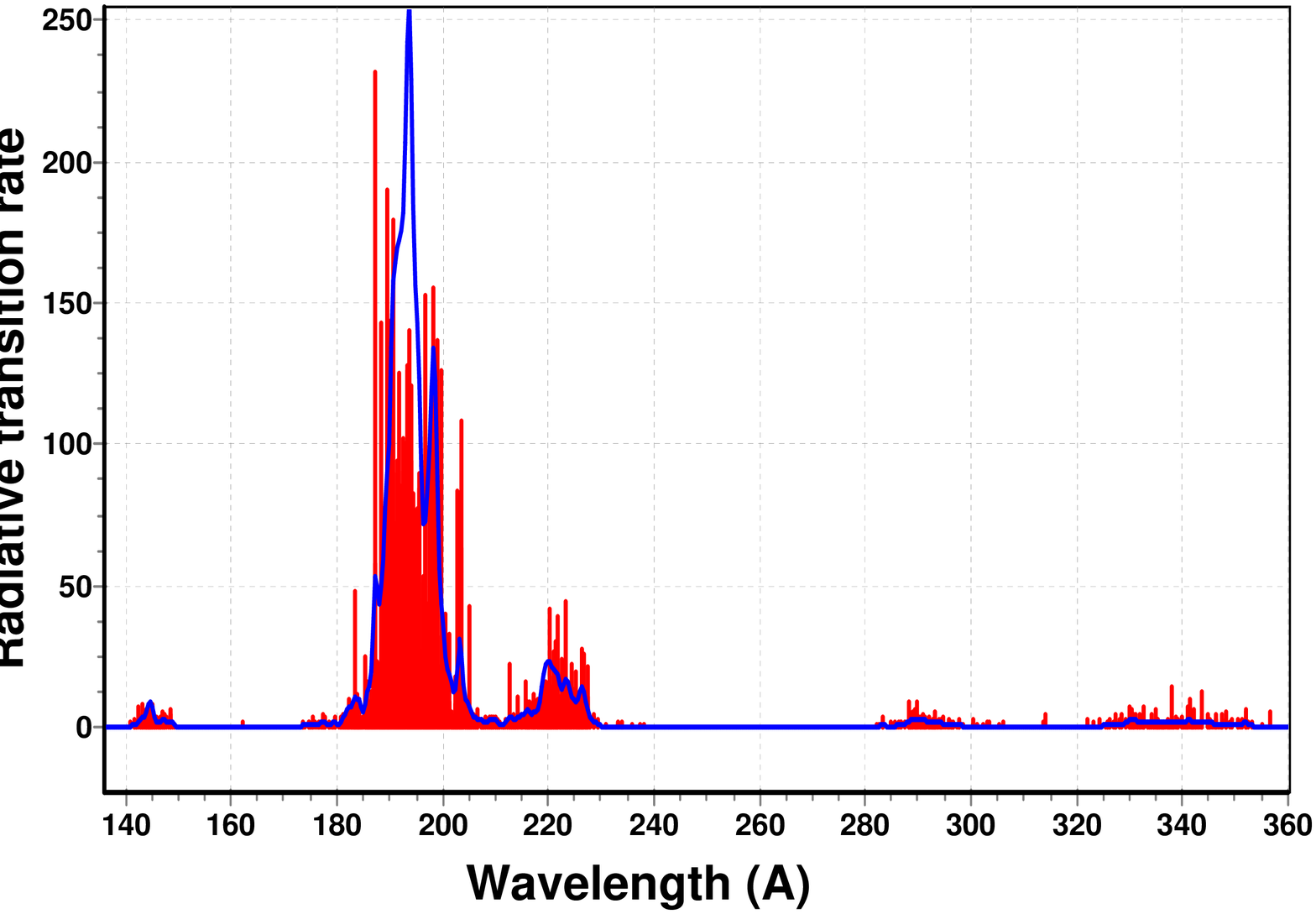}
            \includegraphics[scale=0.35]{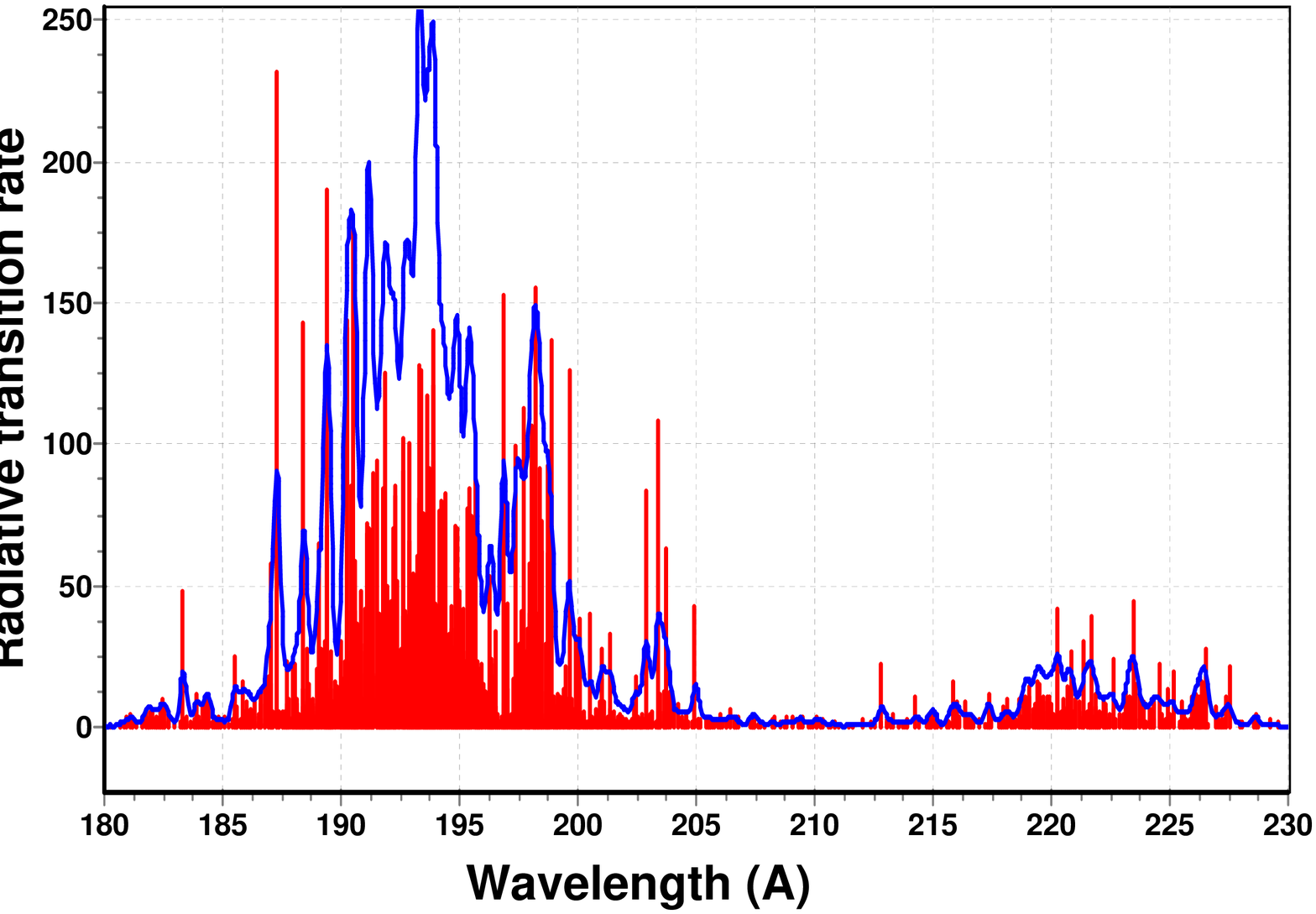}}
\caption{Synthetic spectra (red) for the [$4f^{14}5s$ +
$4f^{13}5s5p$ + $4f^{12}5s5p^2$] $\leftrightarrow$ [$4f^{14}5p$ +
$4f^{13}5s^2$ + $4f^{12}5s^25p$]
 transitions  (red) in
Pm-like Ir$^{16+}$  as a function of wavelength. Promethium is a rare earth element with $Z=61$. A resolving power,
R = $E$/$\Delta E$ = 200 and 600 (left and right) is assumed to
produce a Gaussian profile (blue). The scale in the ordinate is in
units of 10$^{10}$~s$^{-1}$.}
 \label{fig-ir16}
\end{figure*}
\begin{figure*}[tbp]
\centerline{\includegraphics[scale=0.35]{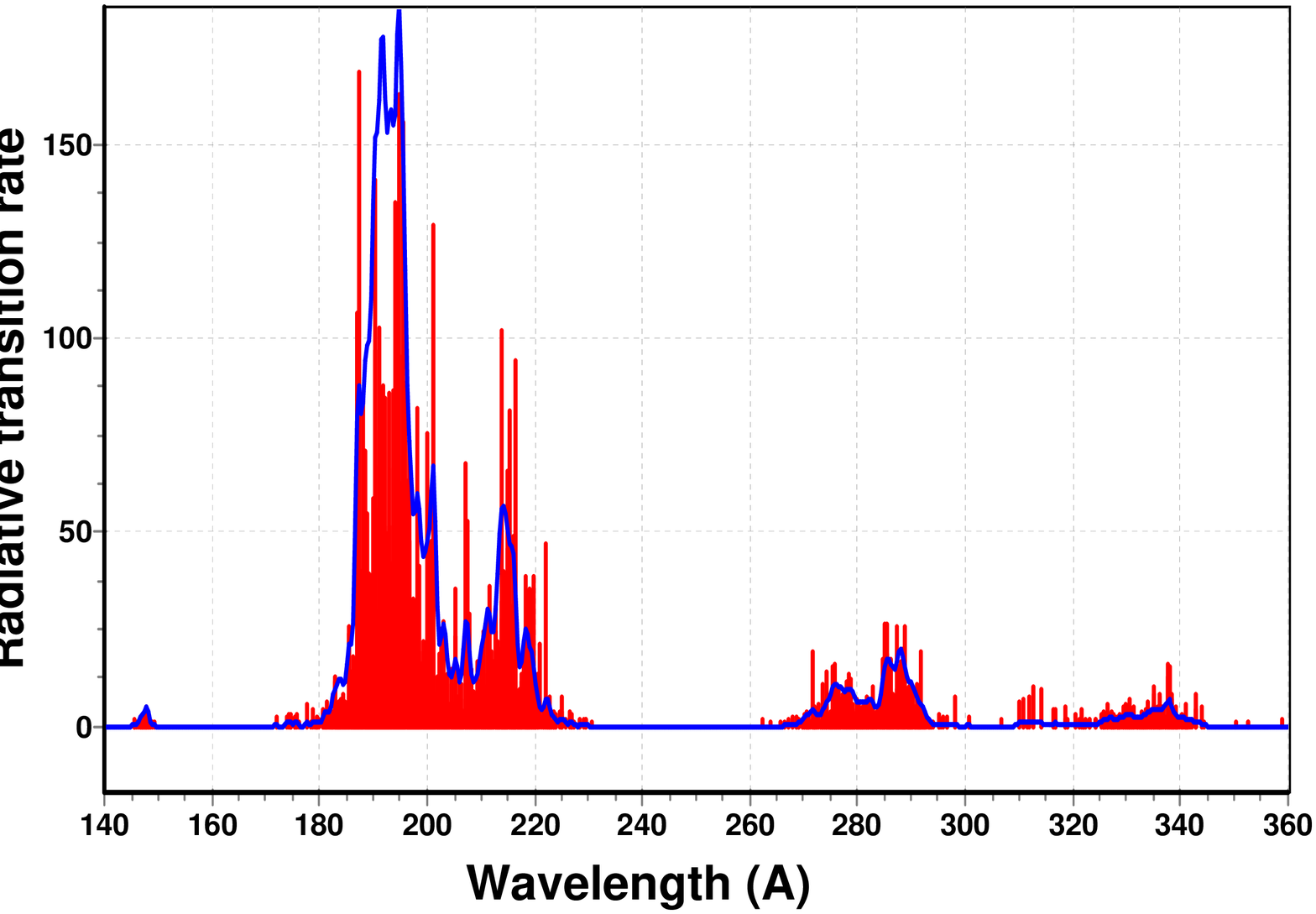}
            \includegraphics[scale=0.35]{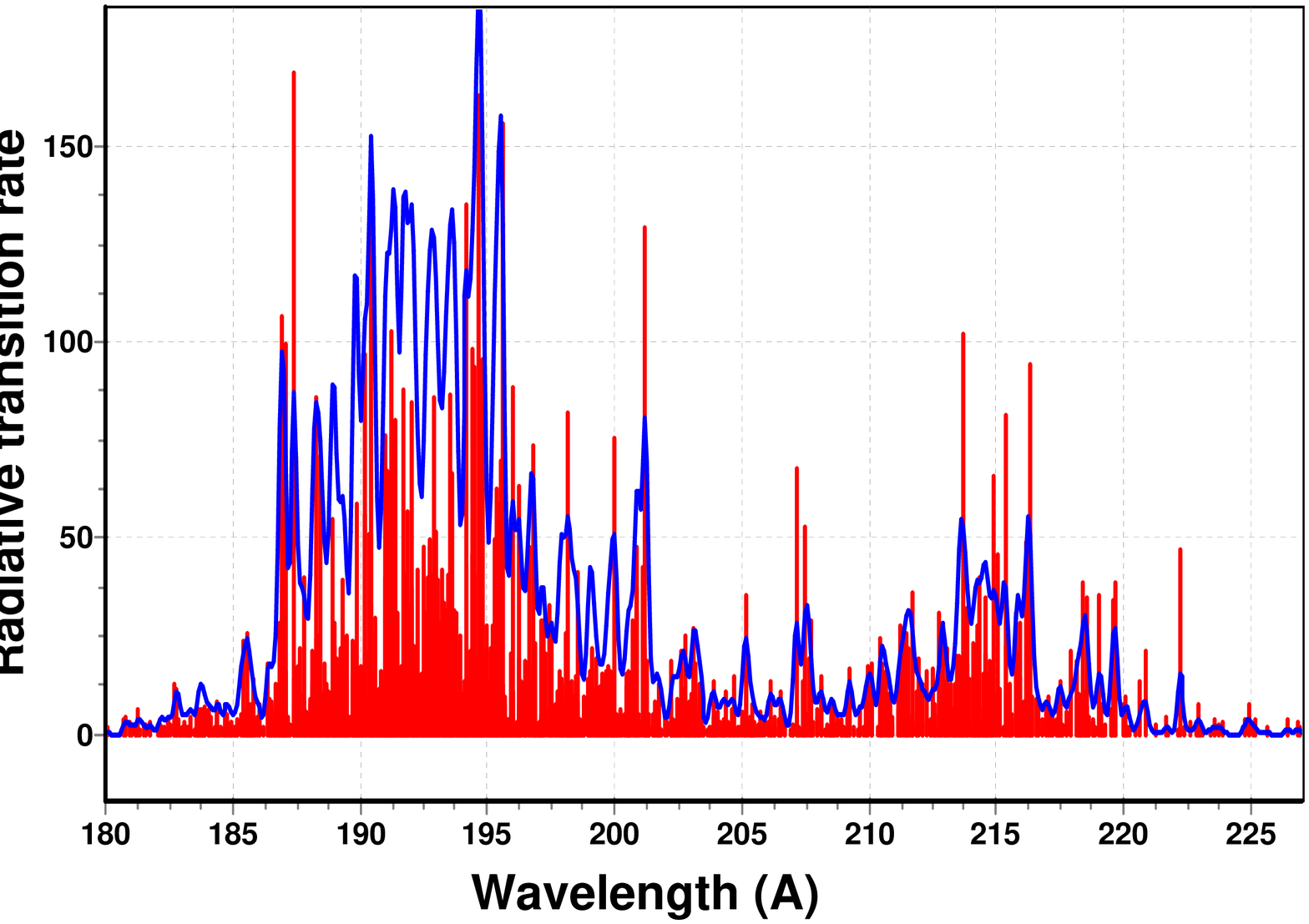}}
\caption{Synthetic spectra (red) for the [$4f^{14}$ + $4f^{13}5p$
+ $4f^{12}5s^2$ + $4f^{12}5p^2$] $\leftrightarrow$ [$4f^{13}5s$ +
$4f^{12}5s5p$]
 transitions  (red)  in
Nd-like Ir$^{17+}$  as a function of wavelength. Neodymium is a rare earth element with $Z=60$.
 A resolving power,
R = $E$/$\Delta E$ = 200 and 800 (left and right) is assumed to
produce a Gaussian profile (blue).
The scale in the ordinate is in
units of 10$^{10}$~s$^{-1}$.}
 \label{fig-ir17}
\end{figure*}
\begin{figure*}[tbp]
\centerline{\includegraphics[scale=0.35]{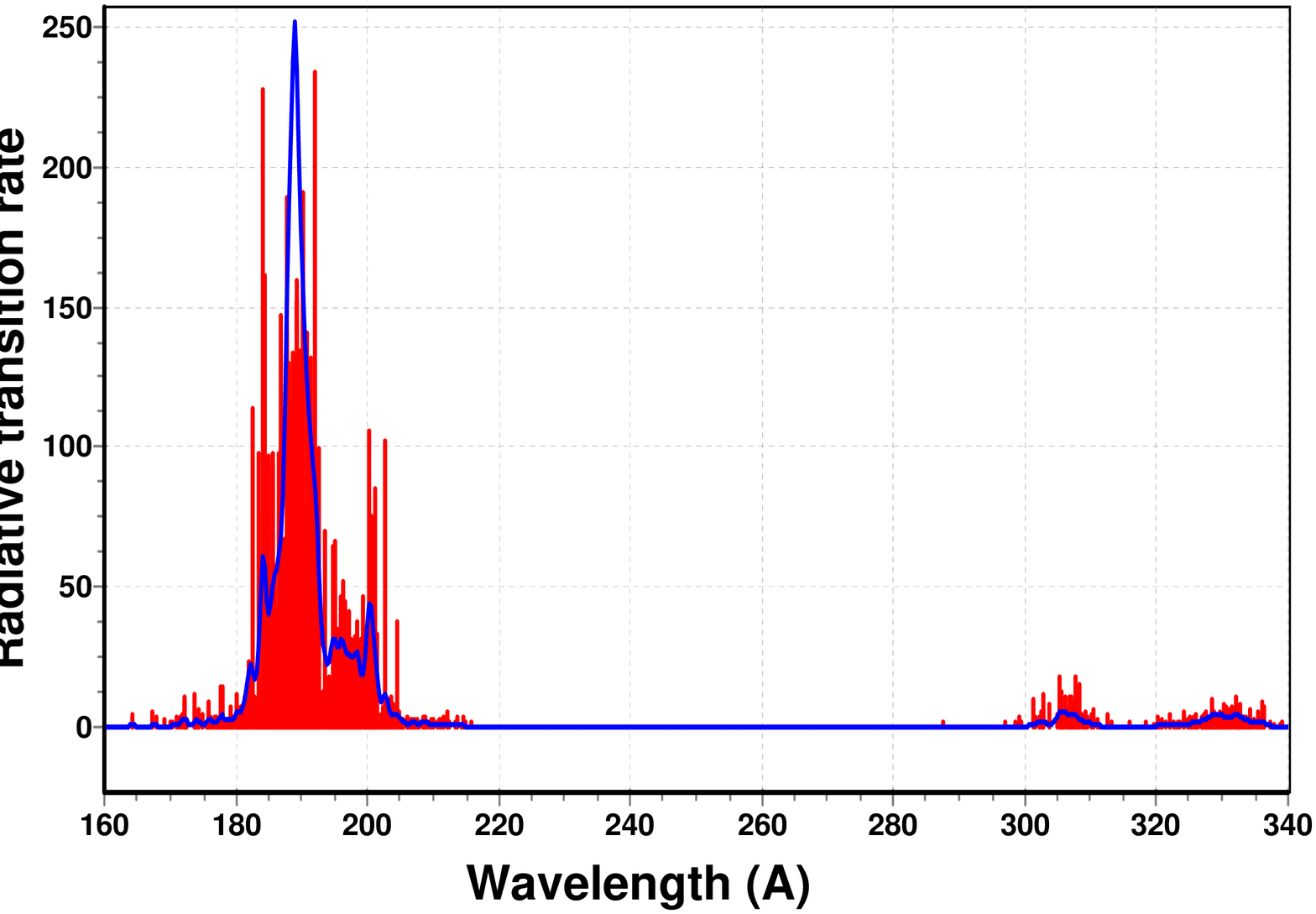}
            \includegraphics[scale=0.35]{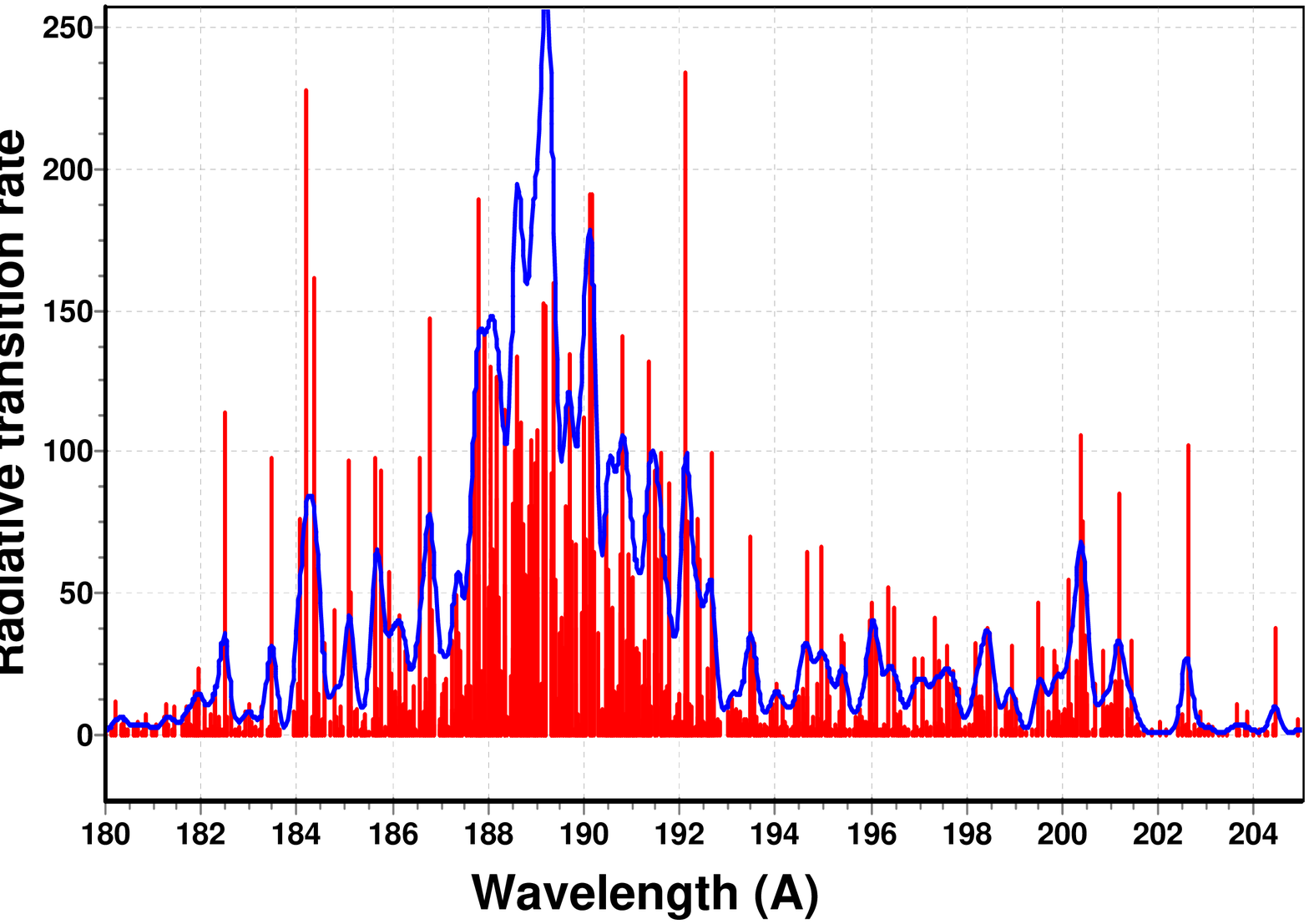}}
\caption{Synthetic spectra (red) for the [$4f^{13}$ + $4f^{12}5p$
+ $4f^{11}5s^2$] $\leftrightarrow$ [$4f^{12}5s$ + $4f^{11}5s5p$]
 transitions  (red)  in
Pr-like Ir$^{18+}$  as a function of wavelength. Praseodymium (Pr) is a rare earth element with $Z=59$.
A resolving power,
R = $E$/$\Delta E$ = 200 and 800 (left and right) is assumed to
produce a Gaussian profile (blue).
The scale in the ordinate is in
units of 10$^{10}$~s$^{-1}$.}
 \label{fig-ir18}
\end{figure*}
 Contributions of the $4f$-core-excited states in determination
 of atomic properties in the
promethium isoelectronic sequence with $Z= 74-92$ were discussed in
Ref.~\cite{pra-13-pm}. Excitation energies, transition rates, and
lifetimes in Pm-like tungsten were evaluated for a large number of states.
The ground state for the Pm-like W$^{13+}$,
Re$^{14+}$, Os$^{15+}$, and Ir$^{16+}$ is $4f^{13}5s^2\
^2F_{7/2}$.
For the next Pm-like ion, Pt$^{17+}$, the $4f^{14}5s\ ^2S_{1/2}$
state becomes the ground state and continues to be the ground state
for higher $Z$ because of the $4f-5s$ level crossing

\section {Energy levels}
Our HFR
calculations include the following set of configurations:
\begin{itemize}
\item Ir$^{16+}$: $4f^{14}5s$, $4f^{14}5p$, $4f^{13}5s^2$, $4f^{13}5p^2$,
$4f^{13}5s5p$, $4f^{12}5s^25p$, and $4f^{12}5s5p^2$.
\item Ir$^{17+}$:
$4f^{14}$,  $4f^{13}5s$, $4f^{13}5p$, $4f^{12}5s^2$,
$4f^{12}5s5p$, and $4f^{12}5p^2$.
\item Ir$^{18+}$: $4f^{13}$,  $4f^{12}5s$, $4f^{12}5p$,
$4f^{11}5s^2$, and  $4f^{11}5s5p$.
\end{itemize}

 In Table~\ref{tab-energy}, we list the
limited number of excitation energies in three iridium ions of interest. The energies are given in 1000 cm$^{-1}$.
The superscipts $a$ and $b$ in this and following tables are seniority numbers used to distinguish levels  
that have the same  electronic configurations,  intermediate and final terms.
We note that the energy differences
 between the doublet $4f^{13}5s^2\
^2F_{J}$ levels and the $4f^{14}5s\ ^2S_{1/2}$ levels in Pm-like
Ir$^{16+}$ are very small, and
the excitation energy of the $4f^{14}5p\ ^2P_{1/2}$ is
larger than the excitation energy of the $4f^{14}5s\ ^2S_{1/2}$ by
a factor of 12. All other levels listed in the third column of
Table~\ref{tab-energy} belong to the $4f^{13}5s5p$ and
$4f^{12}5s^25p$ configurations. First two excited states have transition frequencies
to the ground state in the optical range.
  The energy differences
between the  $4f^{13}5s\ ^3F_{J}$ (with $J$ = 4 and 3) levels and
the $4f^{14}\ ^1S_0$ level in Nd-like Ir$^{17+}$
is 4000-5000~cm$^{-1}$ according to COWAN code. However, the uncertainties in these small energy differences
may be particularly large.
 The excitation
energies  of the two other $4f^{13}5s\ ^3F_{2}$ and $4f^{13}5s\
^1F_{3}$ levels are  larger than the excitation energy of the
 $4f^{13}5s\ ^3F_{3}$ level by a factor of 6-7 and the corresponding transition wavelengths
 to the ground state
  are in optical range. Almost all
other levels listed in the column 6 of Table~\ref{tab-energy}
belong to the $4f^{12}5s^2$ and $4f^{12}5s5p$ configurations. Only
two levels belonging to the  $4f^{13}5p$ configuration have sufficiently low
excitation energies to be included in our list of Nd-like Ir$^{17+}$ levels in
Table~\ref{tab-energy}.

\begin{table*} \caption{\label{tab-br} Wavelengths ($\lambda$ in
\AA), transition rates ($A_r$ in s$^{-1}$),  energies of the lower
and upper level (cm$^{-1}$),  lifetimes ($\tau$ ), and branching
ratios (Branch. ratio) for M1 and E1 transitions in Pm-like Ir$^{16+}$,
Nd-like Ir$^{17+}$, and Pr-like Ir$^{18+}$ ions. }
\begin{ruledtabular}
\begin{tabular}{llllllllllll}
\multicolumn{1}{l}{Conf.} & \multicolumn{1}{l}{Level} &
\multicolumn{1}{l}{Conf.} & \multicolumn{1}{l}{Level} &
\multicolumn{1}{l}{} & \multicolumn{2}{l}{Energies in cm$^{-1}$} &
\multicolumn{1}{l}{$\lambda$} & \multicolumn{1}{l}{$A_r$} &
\multicolumn{1}{l}{Branch.}&
\multicolumn{1}{l}{$\tau$} \\
\multicolumn{2}{l}{Upper Level} & \multicolumn{2}{l}{Lower level} &
 \multicolumn{1}{l}{} &
\multicolumn{1}{l}{Lower} & \multicolumn{1}{l}{Upper} &
\multicolumn{1}{l}{ \AA} & \multicolumn{1}{l}{1/s} &
\multicolumn{1}{l}{ratio}&
\multicolumn{1}{l}{} \\
\hline
\multicolumn{11}{l}{Wavelengths, transition rates,  lifetimes, and branching ratios from M1 and E1 transitions in Pm-like Ir$^{16+}$ ion}\\[0.4pc]\hline
 $ 4f^{13}5s^2  $&$ (^2F) ^2F _{5/2}$&$ 4f^{13}5s^2   $&$(^2F) ^2F _{7/2}$& M1&      0&   25909&   3860& 2.69[+2] &  1.00 & 3.71~ms  \\
 $ 4f^{13}5s5p $&$ (^2F) ^4D _{7/2}$&$ 4f^{13}5s^2    $&$(^2F) ^2F _{7/2}$& E1&      0&   267942&   373& 1.24[+8] &  0.97 & 7.80~ns  \\
 $4f^{12}5s^25p$&$ (^3H)^4G_{11/2}$& $4f^{13}5s^2     $&$(^2F) ^2F_{7/2} $& E2&      0&   276214&   362& 1.19[+2] &  1.00 & 8.37~ms  \\
 $ 4f^{12}5s^25p$&$ (^3H) ^2I_{13/2}$&$ 4f^{12}5s^25p $&$(^3H) ^4G_{11/2}$& M1& 276214&   280512&  23271& 2.15[-1] &  1.00 &  4.65~s \\
 $ 4f^{13}5s5p $&$ (^2F) ^4D _{7/2}$&$ 4f^{13}5s^2    $&$(^2F) ^2F _{7/2}$& E1&      0& 286286&   349& 3.54[+9] &  0.98 &  0.278~ns  \\[0.4pc]

$4f^{12}5s^25p $&$ (^3F) ^4D_{7/2}$ &  $4f^{13}5s5p   $&$(^2F) ^4D _{7/2}$& E1& 267942& 285301&  5761& 2.27[+0] &  0.94 & 0.415~s\\
                &                   &  $4f^{13}5s^2   $&$(^2F) ^2F_{7/2}$ & M1&  0.000& 285301&   351& 5.20[-2] &  0.02 &     \\
                &                   &  $4f^{13}5s^2   $&$(^2F) ^2F_{5/2}$ & M1& 25.909& 285301&   385& 8.49[-2] &  0.04 &     \\[0.4pc]

$4f^{12}5s^25p $&$(^1G) ^2H _{9/2}$&   $4f^{13}5s5p   $&$ (^2F) ^4D _{7/2}$& E1& 267942& 287909& 5008 & 5.75[ 0] &  0.92 &0.161~s\\
                &                  &   $4f^{13}5s^2   $&$ (^2F) ^2F_{5/2} $& E2&  25909& 287909& 382  & 1.80[-1] &  0.03 &      \\
                &                  &   $4f^{12}5s^25p $&$ (^3H) ^4G_{11/2}$& M1& 276214& 287909& 23271& 3.01[-1] &  0.05 &      \\[0.4pc]

 $ 4f^{13}5s5p $&$ (^2F) ^4G _{9/2}$&$ 4f^{13}5s^2    $&$(^2F) ^2F _{7/2}$& E1&      0& 288623&   346& 4.69[+9] &  1.00 &  0.213~ns  \\
 $ 4f^{13}5s5p $&$ (^2F) ^4F _{5/2}$&$ 4f^{13}5s^2    $&$(^2F) ^2F _{7/2}$& E1&      0& 289959&   345& 3.48[+9] &  1.00 &  0.288~ns  \\[0.4pc]
 $ 4f^{13}5s5p $&$ (^2F) ^4G _{5/2}$&$ 4f^{13}5s^2    $&$(^2F) ^2F _{7/2}$& E1&      0& 297128&   337& 1.66[+9] &  0.83 &  0.499~ns\\
                &                   &$ 4f^{13}5s^2    $&$(^2F) ^2F _{5/2}$& E1&  25909& 297128&   369& 3.40[+8] &  0.17 &          \\\hline

\multicolumn{11}{l}{Wavelengths, transition rates,  lifetimes, and branching ratios from M1 and E1 transitions in Nd-like Ir$^{17+}$ ion}\\[0.4pc]\hline
$4f^{13}5s  $&$ (^2F) ^3F_{3}$&  $4f^{13}5s  $&$ (^2F) ^3F_{4}  $&  M1&     0&   4236& 23610&  1.18[+0] &1.00 & 849~ms \\
$4f^{13}5s  $&$ (^2F) ^3F_{2}$&  $4f^{13}5s  $&$ (^2F) ^3F_{3}  $&  M1&  4236&  26174&  4558&  2.26[+2] &1.00 & 4.42~ms \\
$4f^{13}5s  $&$ (^2F) ^1F_{3}$&  $4f^{13}5s  $&$ (^2F) ^3F_{4}  $&  M1&     0&  30606&  3267&  3.05[+2] &1.00 & 3.27~ms \\[0.4pc]
$4f^{12}5s^2$&$ (^3F) ^3F_{4}$&  $4f^{13}5s  $&$ (^2F) ^3F_{3}  $&  E1&  4236&  42199&  2634&  6.51[+1] &0.97 & 14.9~ms \\
                            &&   $4f^{13}5s  $&$ (^2F) ^1F_{3}  $&  E1& 30606&  42199&  8626&  1.79[+0] &0.03 &  \\[0.4pc]
$4f^{12}5s^2 $&$ (^3H) ^3H_{5}$& $4f^{12}5s^2$&$ (^3H) ^3H_{6}  $& M1& 33855&  58261&  4097&  3.80[+2] & 1.00 & 2.63~ms \\[0.4pc]
$4f^{12}5s^2$&$ (^3F) ^3F_{2}$&  $4f^{13}5s  $&$ (^2F) ^3F_{3}  $&  E1&  4236&  63696&  1682&  2.55[+2] & 0.71& 2.79~ms \\
                             & &  $4f^{13}5s  $&$ (^2F) ^1F_{3} $&  E1& 30606&  63696&  3022&  1.04[+2] & 0.29&  \\[0.4pc]
$4f^{12}5s5p$&$ (^3H) ^5G_{6} $&$4f^{12}5s^2$&$ (^3H) ^3H_{6}   $&  E1& 33856& 312026&   359&  2.65[+8] & 0.97& 3.68~ns \\
                              & & $4f^{12}5s^2$&$ (^3H) ^3H_{5} $&  E1& 58261& 312026&   394&  6.72[+6] & 0.03&  \\\hline \multicolumn{11}{l}{Wavelengths, transition rates,  lifetimes, and branching ratios from M1 and E1 transitions in Pr-like Ir$^{18+}$ ion}\\[0.4pc]\hline
$ 4f^{13}    $&$ (^2F) ^2F _{5/2}$& $ 4f^{13}   $&$  (^2F) ^2F _{7/2} $&   M1  &      0&   26442&   3782&   2.86[+2]  &  1.00 &  3.49~ms    \\
$ 4f^{12}5s  $&$ (^3H) ^4H_{11/2}$& $ 4f^{12}5s $&$  (^3H) ^4H _{13/2}$&   M1  &  60142&   67687&  13254&   8.46[+0]  &  1.00 &  118~ms   \\
$ 4f^{12}5s  $&$ (^3F) ^4F _{9/2}$& $ 4f^{13}   $&$  (^2F) ^2F _{7/2}$&    E1  &  0    &   70096&  1427 &   5.92[+2]  &  1.00 &  0.169~ms   \\
 $ 4f^{12}5s  $&$ (^1G) ^2G  _{7/2}$& $ 4f^{12}5s $&$  (^3F) ^4F  _{9/2}$&   M1  &  70096&  74664&  21892&   1.54[+0]  &  1.00 &  650~ms    \\
 $ 4f^{12}5s  $&$ (^3H) ^4H  _{9/2}$& $ 4f^{12}5s $&$  (^3H) ^4H _{11/2}$&   M1  &  67687&  87749&   4985&   2.81[+2]  &  1.00 &  3.55~ms    \\
 $ 4f^{12}5s  $&$ (^3H) ^2H _{11/2}$& $ 4f^{12}5s $&$  (^3H) ^4H _{13/2}$&   M1  &  60142&  89858&   3365&   4.30[+2]  &  1.00 &  2.33~ms    \\
 $ 4f^{12}5s  $&$ (^3F) ^4F  _{5/2}$& $ 4f^{12}5s $&$  (^1G) ^2G  _{7/2}$&   M1  &  74664&  92939&   5472&   4.23[+1]  &  1.00 &  23.7~ms    \\
 $ 4f^{12}5s  $&$ (^3F) ^4F  _{7/2}$& $ 4f^{12}5s $&$  (^3F) ^4F  _{9/2}$&   M1  &  70096&  94507&   4096&   6.23[+1]  &  0.36 &  5.86 ~ms   \\[0.4pc]
               &                    & $ 4f^{12}5s $&$  (^1G) ^2G  _{7/2}$&   M1  &  74664&  94507&   5039&   1.01[+2]  &  0.59 &             \\
 $ 4f^{12}5s  $&$ (^1G) ^2G  _{9/2}$& $ 4f^{12}5s $&$  (^3F) ^4F  _{9/2}$&   M1  &  70096&  98485&   3522&   1.67[+2]  &  0.98 &  5.88 ~ms   \\
   \end{tabular}
\end{ruledtabular}
\end{table*}


\begin{table}
\caption{
 Energies (in ~cm$^{-1}$) in  Nd-like Ir$^{17+}$ ions given relative to
 the   $4f^{13}5s\ ^3F_{4}$ ground states. Comparison of the results calculated by the COWAN and RMBPT codes.
 The difference is given in percent in the last column.}
\begin{ruledtabular}
\begin{tabular}{llllllr}
\multicolumn{1}{c}{Conf.}&
\multicolumn{1}{c}{Level}&
\multicolumn{2}{c}{Energy}&
\multicolumn{2}{c}{Level}&
\multicolumn{1}{c}{DIFF.}\\
\multicolumn{2}{c}{LSJ designations}&
\multicolumn{1}{c}{$E^{\rm COWAN}$}&
\multicolumn{1}{c}{$E^{\rm RMBPT}$}&
\multicolumn{1}{c}{jj designations}&
\multicolumn{1}{c}{J}&
\multicolumn{1}{r}{\%}\\
\hline
   $4f^{13}5s     $&$(^2F) ^3F  _{4}$&        0  &         0&$4f_{7/2}5s_{1/2}$& 4&      \\[0.4pc]
   $4f^{13}5s     $&$(^2F) ^3F  _{3}$&     4236  &      4129&$4f_{5/2}5s_{1/2}$& 3&  2.5 \\
   $4f^{14}       $&$(^1S) ^1S  _{0}$&     5091  &      7055    &$4f^{14}         $& 0&-38.5 \\
   $4f^{13}5s     $&$(^2F) ^3F  _{2}$&    26174  &     25447&$4f_{7/2}5s_{1/2}$& 3&  2.8 \\
   $4f^{13}5s     $&$(^2F) ^1F  _{3}$&    30606  &     30374&$4f_{5/2}5s_{1/2}$& 4&  0.8 \\[0.4pc]

   $4f^{13}5p     $&$(^2F) ^3D  _{3}$&  319802   &   316065&$ 4f_{7/2}5p_{1/2}$& 3&  1.2 \\
   $4f^{13}5p     $&$(^2F) ^3G  _{4}$&  322623   &   319341&$ 4f_{7/2}5p_{1/2}$& 4&  1.0 \\
   $4f^{13}5p     $&$(^2F) ^3G  _{3}$&  346618   &   341890&$ 4f_{5/2}5p_{1/2}$& 3&  1.4 \\
   $4f^{13}5p     $&$(^2F) ^3F  _{2}$&  352344   &   348344&$ 4f_{5/2}5p_{1/2}$& 2&  1.1 \\[0.4pc]

   $4f^{13}5p     $&$(^2F) ^3G  _{5}$&  482729   &   478806&$ 4f_{7/2}5p_{3/2}$&  5&  0.8  \\
   $4f^{13}5p     $&$(^2F) ^3D  _{2}$&  485701   &   481145&$ 4f_{7/2}5p_{3/2}$&  2&  0.9  \\
   $4f^{13}5p     $&$(^2F) ^1F  _{3}$&  491623   &   488430&$ 4f_{7/2}5p_{3/2}$&  3&  0.6  \\
   $4f^{13}5p     $&$(^2F) ^3F  _{4}$&  497788   &   494305&$ 4f_{7/2}5p_{3/2}$&  4&  0.7  \\[0.4pc]

   $4f^{13}5p     $&$(^2F) ^3D  _{1}$&  502142   &   497907&$ 4f_{5/2}5p_{3/2}$&  1& 0.8  \\
   $4f^{13}5p     $&$(^2F) ^3G  _{4}$&  512504   &   507690&$ 4f_{5/2}5p_{3/2}$&  4& 0.9  \\
   $4f^{13}5p     $&$(^2F) ^3F  _{2}$&  519370   &   516116&$ 4f_{5/2}5p_{3/2}$&  2& 0.6  \\
   $4f^{13}5p     $&$(^2F) ^3F  _{3}$&  523054   &   518658&$ 4f_{5/2}5p_{3/2}$&  3& 0.8  \\
 \end{tabular}
\end{ruledtabular}
\label{tab-comp}
\end{table}

\begin{table}
\caption{\label{tab-sp} Wavelengths ($\lambda$ in \AA), weighted
oscillator strengths (gf), weighted transition rates ($gA_r$ in
1/s) for the $4f^{14}5s-4f^{14}5p_{j}$ transitions in Pm-like
Ir$^{16+}$. Comparison of the results evaluated using the first-order
RMBPT1, second-order RMBPT2, and the COWAN codes. Numbers in
brackets represent powers of 10}
\begin{ruledtabular}
\begin{tabular}{llll}
\multicolumn{1}{l}{} &
\multicolumn{1}{l}{RMBPT1} &
\multicolumn{1}{l}{RMBPT2} &
\multicolumn{1}{l}{COWAN}  \\
\hline \multicolumn{1}{c}{}&
\multicolumn{3}{c}{$4f^{14}5s-4f^{14}5p_{1/2}$ transition}\\ $\Delta E$ in cm$^{-1}$&  312299   & 313974     &  309407    \\
 $\lambda$ in \AA      &  320.31    & 318.50     &323.20      \\
 $gf$                 &0.5249       & 0.3809     &  0.3308    \\
 $gA_r$ in 1/s         &3.413[10]   & 2.522[10 ] & 2.112[10]  \\
    \hline \multicolumn{1}{c}{}&
\multicolumn{3}{c}{$4f^{14}5s-4f^{14}5p_{3/2}$ transition}\\
  $\Delta E$ in cm$^{-1}$&    469050    &  472710   &  469871   \\
  $\lambda$ in \AA       &  213.20     &211.55     &212.82  \\
  $gf$                   &   1.6045   & 1.1754      & 1.4855  \\
  $gA_r$ in 1/s          &    2.355[11] &  1.752[11]&  2.187[11]\\
  \end{tabular}
\end{ruledtabular}
\end{table}

The $^2F$ ground state doublet splitting of the
$4f^{13}5s^2$ and $4f^{13}$ configurations in Pm-like Ir$^{16+}$
and Pr-like Ir$^{18+}$ ions differs by only 2\%. The transition from the next excited state
of Ir$^{+18}$ to the ground state
 is already in UV range.  The first 25 low-lying levels of the
Pr-like Ir$^{18+}$ ion  belong to the $4f^{12}5s$ configuration,
while the other 22 levels belong to the
$4f^{11}5s^2$ configuration.

\section {Wavelengths, oscillator strengths, and
transition rates}
In Table~\ref{tab2}, we present selected set of our results
for wavelengths ($\lambda$ in \AA), weighted
oscillator strengths (gf), and weighted transition rates ($gA_r$ in 1/s)
for transitions between the $4f$-core-excited states. Only
transition with the largest values of $gA_r$ ($gA_r >$
10$^{12}$~s$^{-1}$) are given.

We find that the $4f^{12}5s^25p - 4f^{12}5s5p^2$
transitions have the largest values of $gA_r$ for the Pm-like Ir$^{16+}$.
This is expected since these arise from the one-electron electric-dipole $5s-5p$ transitions. The
same type of the transitions are the strongest for the other two ions;
the $4f^{12}5s^2 - 4f^{12}5s5p$ and
$4f^{12}5s5p - 4f^{12}5p^2$ transitions for the Nd-like
Ir$^{17+}$ ion and $4f^{11}5s^2 -
4f^{11}5s5p$ transitions  in Pr-like Ir$^{18+}$ ion, respectively.
The transitions with the largest values of $gA_r$ have wavelengths in the
187~-~200~\AA~ range in Ir$^{16+}$,
187~-~214~\AA~ range in
Ir$^{17+}$,  and 182~-~191~\AA~ range in
 Ir$^{18+}$.

\section{Synthetic spectra}
Synthetic spectra for three Ir highly-charged ions are presented in
Figs.~\ref{fig-ir16},~\ref{fig-ir17}, and~\ref{fig-ir18}, respectively. We
assume that spectral lines have the intensities proportional to
the transition probabilities and are fitted with the Gaussian
profile.

Synthetic spectra displayed in  Figs.~\ref{fig-ir16}-\ref{fig-ir18}
are constructed from the following transitions:
\begin{itemize}
\item Ir$^{16+}$: [$4f^{14}5s$ + $4f^{13}5s5p$ + $4f^{12}5s5p^2$]
$\leftrightarrow$ [$4f^{14}5p$ + $4f^{13}5s^2$ + $4f^{12}5s^25p$],
 \item Ir$^{17+}$:  [$4f^{14}$ + $4f^{13}5p$ + $4f^{12}5s^2$]
$\leftrightarrow$ [$4f^{13}5s$ + $4f^{12}5s5p$],
 \item Ir$^{17+}$: [ $4f^{13}$ + $4f^{12}5p$ +
$4f^{11}5s^2$] $\leftrightarrow$ [ $4f^{12}5s$ + $4f^{11}5s5p$].
\end{itemize}

Every spectrum on the left panel of the figure includes  about 2000 transitions with the
values of $gA_r >10^{10}$~s$^{-1}$.
The synthetic spectra on the right
panel include lines with the largest $gA_r$ values. For example, the
spectrum of  Ir$^{16+}$ displayed on the left panel of
Fig.~\ref{fig-ir16} includes the spectral region
 140\AA~ - 360\AA~. In the right panel of Fig.~\ref{fig-ir16}, we limit this region to 180\AA~ - 230\AA~ by
 neglecting the part of spectra with small intensity with the $A_r$
value less than 10 in units of 10$^{10}$s$^{-1}$.  The
similar procedure was used for the other synthetic spectra.

Comparison of  spectra at the right panels of of
Figs.~\ref{fig-ir16},~\ref{fig-ir17}, and~\ref{fig-ir18}
 shows the similarities including the main peak,
some strong lines separate from the main peak and the second wide
peak of small intensity. All strong lines are listed in
Table~\ref{tab2} and Supplemental Material \cite{suppl}.

We note that present synthetic spectra  do not take into account the population of states, and are meant to illustrate the distribution on the 
strongest lines. In  typical EBIT conditions many of the states are barely populated, making the spectra significantly less dense ~\cite{n1,n2}.

\section {Multipole transitions, branching ratios, and lifetimes}
Wavelengths, transition rates,  energies of the lower and upper
level, lifetimes, and branching ratios  from M1 and E1 transitions
in Nd-, Pm-, and Pr-like Ir ions are presented in
Table~\ref{tab-br}.
 In order to determine the lifetimes
listed in the last columns of Table~\ref{tab-br},
we sum over all possible radiative transitions.
The value of branching ratios for the particular transition is
determined as a ratio of the respective $A_r$ values and
the sum of all possible radiative transition rates that are used to determine
the lifetimes.  The number of contributing transitions increases
significantly for higher levels. To save space, we only included
the transitions that give the largest contributions to the lifetimes, and list additional
transitions in the Supplemental Material \cite{suppl}.

We use atomic units (a.u.) to express all transition matrix
elements throughout this section: the numerical values of the
elementary
 charge, $e$, the reduced Planck constant, $\hbar = h/2
\pi$, and the electron mass, $m_e$, are set equal to 1. The atomic
unit for electric-dipole matrix element is  $ea_0$, where $a_0$ is
the Bohr radius.

The E1 and M1 transition probabilities $A_r$ (s$^{-1}$) are
obtained in terms of line strengths $S$~(a.u.) and wavelengths
$\lambda$~(\AA) as
\begin{eqnarray}
A(E1)&=&2.02613\times 10^{18}\frac{S(E1)}{(2J+1)\lambda ^{3}},\\
A(M1)&=&2.69735\times 10^{13}\frac{S(M1)}{(2J+1)\lambda ^{3}}.
\end{eqnarray}
The line strengths $S(E1)$ and $S(M1)$ are obtained as squares of
the corresponding E1 and M1  matrix elements.

\begin{table*}
\label{taba}
\caption{ Comparison of the energies (in ~cm$^{-1}$) in  Nd-like Ir$^{17+}$ ion given relative to
 the   $4f^{13}5s\ ^3F_{4}$ ground state obtained in different approximations \cite{prl-11,prl-15}. }
\begin{ruledtabular}
\begin{tabular}{rrrrrrrr}
\multicolumn{1}{c}{Conf.}&
\multicolumn{1}{c}{Level}&
\multicolumn{2}{c}{Energies in cm$^{-1}$}&
\multicolumn{1}{c}{Diff.}&
\multicolumn{2}{c}{Energies in cm$^{-1}$}&
\multicolumn{1}{c}{Diff.}\\
\multicolumn{2}{c}{}&
\multicolumn{1}{c}{ COWAN}&
\multicolumn{1}{c}{CI~\cite{prl-11}}&
\multicolumn{1}{c}{\%}&
\multicolumn{1}{c}{FSCC~\cite{prl-15}}&
\multicolumn{1}{c}{CIDFS~\cite{prl-15}}&
\multicolumn{1}{c}{\%}\\
\hline
   $4f^{13}5s     $&$(^2F) ^3F  _{4}$&        0&      0    &     &          &         &     \\
   $4f^{13}5s     $&$(^2F) ^3F  _{3}$&     4236&    4838   &  12\% &    4662 &    4872 &     4\%\\
   $4f^{13}5s     $&$(^2F) ^3F  _{2}$&    26174&    26272  &   0\% &   25156  &    25044&    0\% \\
   $4f^{13}5s     $&$(^2F) ^1F  _{3}$&    30606&    31492  &   3\% &   30197  &    30552&    1\% \\[0.4pc]
   $4f^{14}       $&$(^1S) ^1S  _{0}$&     5091&    5055   &  -1\% &   13599  &    7025 &  -94\% \\[0.4pc]
   $4f^{12}5s^2   $&$(^3H) ^3H  _{6}$&    33856&    35285  &   4\% &   24221  &   29367 &   18\% \\
   $4f^{12}5s^2   $&$(^3F) ^3F  _{4}$&    42199&    45214  &   7\% &   33545  &   38295 &   12\% \\
   $4f^{12}5s^2   $&$(^3H) ^3H  _{5}$&    58261&    59727  &   2\% &   47683  &   52668 &    9\% \\
   $4f^{12}5s^2   $&$(^3F) ^3F  _{2}$&    63696&    68538  &   7\% &   55007  &   60322 &    9\% \\
   $4f^{12}5s^2   $&$(^3H) ^1G  _{4}$&    66296&    68885  &   4\% &   56217  &   60943 &    8\% \\
   $4f^{12}5s^2   $&$(^3F) ^3F  _{3}$&    68886&    71917  &   4\% &   58806  &   63847 &    8\% \\
   $4f^{12}5s^2   $&$(^3H) ^3H  _{4}$&    89455&    92224  &   3\% &   78534  &   82954 &    5\% \\
   $4f^{12}5s^2   $&$(^3P) ^1D  _{2}$&    91765&    98067  &   6\% &   82422  &   88261 &    7\% \\
   $4f^{12}5s^2   $&$(^1I) ^1I  _{6}$&   101537&   110065  &   8\% &   93867  &  101844 &    8\% \\
   $4f^{12}5s^2   $&$(^3P) ^3P  _{0}$&   101073&   110717  &   9\% &   94012  &   99617 &    6\% \\
   $4f^{12}5s^2   $&$(^3P) ^3P  _{1}$&   107843&   116372  &   7\% &   99416  &  105989 &    6\% \\
   $4f^{12}5s^2   $&$(^1D) ^3P  _{2}$&   117322&           &     &  107489  &  113272 &    5\%\\
   $4f^{12}5s^2   $&$(^1S) ^1S  _{0}$&   178055&           &     &  174893  &  185757 &    6\%\\
 \end{tabular}
\end{ruledtabular}
\end{table*}

In Table~\ref{tab-br}, we include results for 8 selected
electric-dipole and magnetic-mulipole transitions that
are the most important for the evaluation of the corresponding
lifetimes in Pm-like Ir$^{16+}$ ion.
Results for 48 transitions contributing to the lifetimes of 32 lowest levels are
listed in the Supplemental Material \cite{suppl}.
Transitions with small branching ratios are omitted from the table.

The second excited state, $4f^{14}5s\ ^2S_{1/2}$, is metastable with extremely long lifetime
since the strongest possible decay channels are
 electric-octupole (E3) transition to the ground state, which is in optical range,  and
magnetic-quadrupole (M2) transition for the first excited level, $4f^{13}5s^2\
^2F_{5/2}$. These transitions are too weak to be estimated with the COWAN code, so we described such cases in text
but omit from the Table~\ref{tab-br}. There is only one such level for Ir$^{16+}$ and Ir$^{16+}$ ions and two levels
for Ir$^{17+}$ ion.

The lifetime of the
first excited $4f^{13}5s^2\ ^2F_{5/2}$ state
is equal to 3.71~ms with $A_r$
equal to 269~s$^{-1}$.
The lifetime of the third excited state is very short, 7.8 ns due to E1 allowed
$4f^{13}5s^{2}\  ^2F_{7/2} - 4f^{13}5s5p\ ^4D_{7/2}$
transition.
The $4f^{12}5s^25p\
^2I_{13/2}$ excited state is metastable with
4.65~s lifetime since  the strongest transition is
$4f^{12}5s^25p\ ^4G_{11/2} -   4f^{12}5s^25p\
^2I_{13/2}$  with small transition energy,
4298~cm$^{-1}$.

 While some of the E1 transitions listed in Table~\ref{tab-br} are strong
 allowed $5s-5p$ transitions, a number of E1 transitions are very weak since these
 are forbidden transitions with non-zero amplitude due to configuration mixing.
For example, the $A_r$ value of the $ 4f^{13}5s^2\
^2F_{7/2}$ - $4f^{13}5s5p \   ^4G _{9/2}$ transition is lager
by nine orders of  magnitude than $4f^{13}5s5p\  ^4D_{7/2}$ -
$4f^{12}5s^25p\ ^4D_{7/2}$ transition. In the first case, we have one-electron
E1-allowed $5s-5p$ transition, while the second case is a strongly
forbidden $4f-5s$ transition. The non-zero value of the $4f^{13}5s5p\
 ^4D_{7/2}$ - $4f^{12}5s^25p\ ^4D_{7/2}$ matrix element is
due to mixing configurations.

In Table ~\ref{tab-br}, we include results for 10 selected
electric-dipole and magnetic-mulipole  transitions that
are the most important for the evaluation of the corresponding Ir$^{17+}$
lifetimes. Results for 33 transitions contributing to the lifetimes of 13 lowest levels are
listed in the Supplemental Material \cite{suppl}.
 We also evaluated
electric-quadrupole transitions, however, their contributions to lifetimes of levels
given in Table~\ref{tab-br} is negligible.  Four of the transitions
listed in Table~\ref{tab-br} are M1 transitions between the states inside of the
$4f^{13}5s$ or $4f^{12}5s^2$ configurations. While
the $4f^{13}5s - 4f^{12}5s^2$ transitions are E1-forbidden,  their transition rates
 are non-zero due to mixing of the $4f^{13}5s$
and the  $4f^{12}5s5p$ configurations. The lifetimes of most levels given in
Table ~\ref{tab-br} are relatively long, since there are no allowed E1
transitions that may contribute to these lifetimes until the first level containing the $5p$ electron.
 The resulting lifetime of the $4f^{12}5s5p\  ^5G_{6}$ level is very short,
  3.68~ns due to the $4f^{12}5s^{2} - 4f^{12}5s5p$ E1
 transitions.

Two Ir$^{17+}$ levels, $4f^{14}\ ^1S_{0}$ and  $4f^{12}5s^2\  ^3H_{6}$, are extremely long lived making then potential candidates for the atomic clock
 scheme implementation.
 There is only   extremely weak electric-octupole transition to the $4f^{13}5s\ (^2F) ^3F_{3}$
 state contributing to $^1S_{0}$ lifetime.
 The $4f^{12}5s^2\ ^3H_{6}$ lifetime is determined
by the two the E3 and one M2 channels:
 $4f^{13}5s\  ^3F_{3}$ -  $4f^{12}5s^2\ ^3H_{6}$, $4f^{13}5s\  ^1F_{3}$ -  $4f^{12}5s^2\  ^3H_{6}$, and $4f^{13}5s\ ^3F_{4}$ -  $4f^{12}5s^2\  ^3H_{6}$.

In Table~\ref{tab-br}, we include results for 10 selected
electric-dipole  and magnetic-multipole  transitions that
are most important for the evaluation of the corresponding
lifetimes in Pr-like Ir$^{18+}$ ion. Results for 51 transitions contributing to the lifetimes of 29 lowest levels are
listed in the Supplemental Material \cite{suppl}.
 Almost all transitions in
Table~\ref{tab-br} are magnetic-dipole transitions. Only last
two lines include E1 forbidden transitions between
$4f^{12}5s$ and $4f^{11}5s^2$ configurations. List of levels for
the  Pr-like Ir$^{18+}$ ion in Table~\ref{tab-energy} covers the
energy interval from the ground state up to 274768~cm$^{-1}$. The
first level with $5p$ electron ($4f^{12}5p\  ^4G_{11/2}$
level) has energy equal to 387658~cm$^{-1}$. As a result, there are no
electric-dipole allowed transitions in Table~\ref{tab-br}.

The only extremely long-lived metastable level in Ir$^{+18}$ ion
 is  $4f^{12}5s\  ^4H_{13/2}$ third excited state with
$E=60142$~cm$^{-1}$.  This state can only decay via  very weak
electric-octupole transition to the ground state, which is in UV range.

 The next longest lifetime, 8100~s, is
for the $4f^{12}5s\  ^2I_{11/2}$ level due to very small,
205~cm$^{-1}$, energy difference in the  $4f^{12}5s\
^2I_{13/2} - 4f^{12}5s\ ^2I_{11/2}$ transition.
 The $4f^{11}5s^2\ ^2I_{13/2}$ level has short lifetime
 due to contribution
of two E1 transitions  via level mixing with branching ratios equal to 62\% and 38\%.

Dominant contribution to the $4f^{12}5s  ^4F _{9/2}$ lifetime is from E1-forbidden transition to
 the $4f^{13}$~$^2F_{7/2}$  state which is non-zero due to the 5\% mixing between $4f^{13}$ and $4f^{12}
5p$~  configurations.

We did not find any theoretical or
experimental results to compare with our $A_{r}$ and $\tau$ values
for the low-lying states listed in
 Table~\ref{tab-br}.

\begin{table*}
\caption{
Comparison of the present results with theoretical \cite{prl-11,prl-15}  and  experimental \cite{prl-15}
transition energies (in ~cm$^{-1}$) for M1 transitions in  Nd-like Ir$^{17+}$ ion. Differences between experimental
and theoretical results are given in per cent for all theoretical values.}
\label{tabb}
 \begin{ruledtabular}
\begin{tabular}{rrrrrrrrrrrr}
\multicolumn{1}{c}{Conf.}&
\multicolumn{1}{c}{Transition}&
\multicolumn{1}{c}{Energy}&
\multicolumn{1}{c}{Energy}&
\multicolumn{1}{c}{Diff.}&
\multicolumn{1}{c}{Energy}&
\multicolumn{1}{c}{Diff.}&
\multicolumn{1}{c}{Energy}&
\multicolumn{1}{c}{Diff.}&
\multicolumn{1}{c}{Energy}&
\multicolumn{1}{c}{Diff.}\\
\multicolumn{1}{c}{}& \multicolumn{1}{c}{}&
\multicolumn{1}{c}{Expt.~\cite{prl-15}}&
\multicolumn{1}{c}{COWAN}&
\multicolumn{1}{c}{\%}&
\multicolumn{1}{c}{CI~\cite{prl-11}}&
\multicolumn{1}{c}{\%}&
\multicolumn{1}{c}{FSCC~\cite{prl-15}}&
\multicolumn{1}{c}{\%}&
\multicolumn{1}{c}{CIDFS~\cite{prl-15}}&
\multicolumn{1}{c}{\%}\\
\hline
$4f^{13}5s   $&$ ^3F_{2}\ -\ ^3F_{3}$&    20710.83&    21938&  -6\%& 21434&   -3\%&   20494&    1\%& 20172&  3\%\\
$4f^{12}5s^2 $&$ ^3H_{4}\ -\ ^1G_{4}$&    22430.03&    23159&  -3\%& 23339&   -4\%&   22317&    1\%& 22011&  2\%\\
$4f^{12}5s^2 $&$ ^1G_{4}\ -\ ^3F_{4}$&    22948.54&    24097&  -5\%& 23671&   -3\%&   22672&    1\%& 22648&  1\%\\
$4f^{12}5s^2 $&$ ^1D_{2}\ -\ ^3F_{3}$&    23162.84&    22879&   1\%& 26150&  -13\%&   23616&   -2\%& 24414& -5\%\\
$4f^{12}5s^2 $&$ ^3H_{5}\ -\ ^3H_{6}$&    23639.87&    24405&  -3\%& 24442&   -3\%&   23462&    1\%& 23301&  1\%\\
$4f^{12}5s^2 $&$ ^3F_{3}\ -\ ^3F_{4}$&    25514.56&    26687&  -5\%& 26703&   -5\%&   25261&    1\%& 25552&  0\%\\
$4f^{12}5s^2 $&$ ^1D_{2}\ -\ ^3F_{2}$&    27387.06&    28069&  -2\%& 29529&   -8\%&   27415&    0\%& 27939& -2\%\\
$4f^{13}5s   $&$ ^1F_{3}\ -\ ^3F_{4}$&    30358.45&    30606&  -1\%& 31492&   -4\%&   30197&    1\%& 30552& -1\%\\
$4f^{12}5s^2 $&$ ^3H_{4}\ -\ ^3H_{5}$&    30797.17&    31194&  -1\%& 32497&   -6\%&   30851&    0\%& 30286&  2\%\\
 \end{tabular}
\end{ruledtabular}
\label{tab-br-pr}
\end{table*}

\section{Comparisons, Uncertainty estimates, and Conclusion }

In Table~\ref{tab-comp}, we compare energies  in Nd-like
Ir$^{17+}$ ion calculated using the COWAN and RMBPT codes. Details of
the RMBPT code for the hole-particle systems were presented by
Safronova et. al.~\cite{ni-00}.
We use a complete
set of Dirac-Fock (DF) wave functions on a nonlinear grid generated using
B-splines \cite{Bspline} constrained to a spherical cavity. The
basis set consists of 50 splines of order 9 for each value of the
relativistic angular quantum number $\kappa$. The starting point for the RMBPT code
is the \textit{frozen core} DF approximation.
 We use second-order RMBPT to
determine energies of  the $4f^{13}5s$ states relatively to the
$4f^{14}$ state in Nd-like Ir$^{17+}$ ion.

The $jj$ designation
($4f_{7/2}5s_{1/2}$ for example) listed in
Table~\ref{tab-comp} means that we consider the system with
$4f_{7/2}$ hole  in $4f^{14}$  core and  the $5s_{1/2}$ particle.
The $LS$ designation are given in columns ``1'' and ``2'' of
Table~\ref{tab-comp}. The results with the same
configuration and $J$ are compared. The difference of RMBPT and COWAN data  given in
the last column of Table~\ref{tab-comp} is less than  1\% for most levels.
The difference exceeds 3\% only for $4f^{14}$ level.
This is the only level in the table that has a different number of $4f$
electrons in comparison with the ground state. The correlation correction is very large
for the $4f$ state leading to this difference.
For example, the calculation of the $4f_{7/2}5s_{1/2}$~$J=4$ ground state energy gives
5447~cm$^{-1}$ in lowest order DF approximation, while the second order value is
almost five times larger,  -23997~cm$^{-1}$. Adding the 4577~cm$^{-1}$ QED and 6888~cm$^{-1}$
Breit corrections gives -7055~cm$^{-1}$ relative to the frozen-core $4f^{14}$ level.

In Table~\ref{tab-sp}, we compare wavelengths, weighted oscillator
strengths, and weighted transition rates  for the
$4f^{14}5s-4f^{14}5p$ transitions in Pm-like Ir$^{16+}$
 evaluated using the first-order RMBPT1,
 second-order RMBPT2~\cite{ni-00}, and the COWAN codes. COWAN code energy values
 are  in somewhat better agreement with the first-order RMBPT results. The difference of
 RMBPT and COWAN code energies and wavelength is small, 0.18~\% - 1.5~\%,
for wavelengths and energies of the $4f^{14}5s-4f^{14}5p$.

The differences are larger, 7~\% - 27~\%, for the
oscillator strength and transition rates. We find large effects of the second order corrections
for these properties leading to larger differences in the results.
In summary, the second-order correlation
corrections are very important for accurate determination of the
oscillator strengths and transition rates.

Energies (in ~cm$^{-1}$) in  Nd-like Ir$^{17+}$ ion relative to
 the   $4f^{13}5s\ ^3F_{4}$ ground state obtained by the COWAN
 code are compared with results from Ref.~\cite{prl-11} and ~\cite{prl-15} in Table~\ref{taba}.
 Results in \cite{prl-11} were obtained using configuration-interaction (CI) approach.
Results in \cite{prl-15} were obtained using multireference Fock space
coupled cluster (FSCC) method and
configuration-interaction-Dirac-Fock-Sturmian (CIDFS) method. COWAN results were also presented in ~\cite{prl-15}.
  The
largest difference (12\%) between the COWAN and CI~\cite{prl-11}
results is for the first excited state. For other levels, the
difference is about 3-4\%. The largest difference (94\%) between
results in Ref.~\cite{prl-15}, evaluated by FSCC and CIDFS methods
 are for the $4f^{14}\ ^1S_0$ level. For other levels, the
difference is about 5-9\%.

Comparison of the present results with theoretical \cite{prl-11,prl-15}  and  experimental \cite{prl-15}
transition energies (in ~cm$^{-1}$) for M1 transitions in  Nd-like Ir$^{17+}$ ion are given in Table~\ref{tabb}.
 In \cite{prl-15}, improved accuracy measurements
were performed for the Ir$^{17+}$ transitions by repeating a
calibration-measurement-calibration cycle five times at one
grating position. Each cycle takes 15-30 min. The transition
energy was determined by averaging the line centroids in typically
30 of such acquired spectra.
Results are given for the M1 transitions between the states of the $4f^{13}5s$ and  $4f^{12}5s^2$ configurations.
 The differences between the experimental results and theoretical values obtained by the COWAN code,
  CI \cite{prl-11}, FSCC and CIDFS~\cite{prl-15} are given in percent. The largest difference between the experimental values \cite{prl-15}
 and the COWAN code results is 6\%, while the difference is twice as large for CI values~\cite{prl-11}.
 The differences between the FSCC~\cite{prl-15} and CIDFS~\cite{prl-15} results  and experiment \cite{prl-15} are 0-2\% and 0-5\%, respectively.

In summary, we carried out  a systematic study of excitation energies, wavelengths, oscillator strengths, and
transition rates in Ir$^{16+}$,
Ir$^{17+}$, and Ir$^{18+}$ ions.
Synthetic spectra are constructed for all three ions, with the strongest line
regions presented in more detailed on a separate panel.   Metastable states and
importance of the M1 transitions for the determination of the lifetimes are discussed.
Comparison of the energy values with other theoretical predictions is given.
Transition wavelengths are compared with the
experiment ~\cite{prl-15} and other theory.
 We did not find any other theoretical
results of oscillator strength and transitions rates in the ions considered in the present study.
We believe that our predictions will be useful for planning and analyzing future
experiments with highly-charged ions.
\section*{Acknowledgements}

We thank H. Bekker, A. Windberger,  and
J. R. Crespo L\'opez-Urrutia for discussions and comments on the paper. 
 M. S. S. thanks the School
of Physics at UNSW, Sydney, Australia and MPIK, Heidelberg, Germany for hospitality and
acknowledges support from the Gordon Godfrey
Fellowship program, UNSW. This work was supported in part by U. S. NSF Grant
No. PHY-1404156 and the Australian Research Council.

\end{document}